# 2D Solitons in $\mathcal{PT}$-symmetric photonic lattices

André L. M. Muniz[1], Martin Wimmer[1], Arstan Bisianov[1], Pawel S. Jung[2], Demetrios N. Christodoulides[2], Roberto Morandotti[3,4,5] and Ulf Peschel[1]

1. Abbe Center of Photonics, Friedrich Schiller University Jena, Max-Wien-Platz 1, 07743 Jena, Germany
2. CREOL, College of Optics and Photonics, University of Central Florida, Orlando, Florida 32816–2700, USA
3. INRS EMT, 1650 Blvd Lionel Boulet, Varennes, PQ J3X 1S2, Canada
4. ITMO University, St. Petersburg, Russia
5. Institute of Fundamental and Frontier Sciences, University of Electronic Science and Technology of China, Chengdu 610054, China

**Abstract**

Parity-time ($\mathcal{PT}$) symmetry has attracted a lot of attention since the concept of pseudo-Hermitian dynamics of open quantum systems was first demonstrated two decades ago. Contrary to their Hermitian counterparts, non-conservative environments *a priori* do not show real energy eigenvalues and unitary evolution. However, if $\mathcal{PT}$-symmetry requirements are satisfied, even dissipative systems can exhibit real energy eigenvalues, thus ensuring energy conservation in the temporal average. In optics, $\mathcal{PT}$-symmetry can be readily introduced by incorporating, in a balanced way, regions having optical gain and loss. However, all optical realizations have been restricted so far to a single transverse dimension (1D) such as optical waveguide arrays. In many cases only losses were modulated relying on a scaling argument being valid for linear systems only. Both restrictions crucially limit potential applications. Here, we present an experimental platform for investigating the interplay of $\mathcal{PT}$-symmetry and nonlinearity in two dimensions (2D) and observe nonlinear localization and soliton formation. Contrary to the typical dissipative solitons, we find a one-parametric family of solitons which exhibit properties similar to its conservative counterpart. In the limit of high optical power, the solitons collapse on a discrete network and give rise to an amplified, self-accelerating field.

**Manuscript**

Light is by far the most important information carrier in modern society, but has also started to influence material processing in industry and is expected to become the basis of future computing schemes. In all these applications two fundamental drawbacks are experienced: absorption of photons and spreading of initially confined wave packets. Two major roads are followed to fight these obstacles: amplification to beat losses and soliton formation, that nonlinear action compensates for linear dispersive forces. Being rather successful individually, a combination of both approaches is not trivial. A restoration of conservative features requires delicate adjustment, which in the case of failure results in a decay of injected signals or in their explosive growth which is finally limited by saturation of amplification only. In the latter case so-called dissipative solitons are formed, the nature of which reflects the input field. In addition low amplitude noise is often amplified around dissipative solitons or they even require a background to exist. In contrast so-called PT-symmetric systems in which lossy and amplifying sections are combined with phase modulation offer the unique possibility to restore a quasi-conservative situation. PT-symmetric systems also offer a öot of other interesting features as e.g. a phase transition for growing gain modulation or unidirectional invisibility. Combined with nonlinearity one expects an even richer spectrum of phenomena. It has been shown by various simulations that whole soliton families exist in one- and two-dimensional PT symmetric systems with Kerr nonlinearity. However, experimental verifications have up to date only be obtained for one genuine transverse dimension only. It would be beneficial to realize also a two dimensional nonlinear version of a PT symmetric system as the properties of solitons depend on the dimension critically. This is best understood for the Nonlinear Schrödinger Equation, which is lossless and supports bright solitons for focusing Kerr nonlinearity. Their energy is inverse to their width in one-dimension, but constant for two-dimensional systems. In the latter case a collapse of the field distribution may occur as a contraction does not require additional power. Such properties are nearly reproduced in the PT symmetric case, except the collapse, which can be arrested by the inherent discreteness of PT symmetric systems due to the internal gain and index modulation. Here we work with a fiber-based system as it provides easy access to gain, loss and phase modulation based on standard telecommunication equipment. To realize the attractive features of two-dimensional solitons, we make use of the newly developed concept of synthetic dimensions. By combining short- and long range interaction, we mimic an effective discrete two-dimensional lattice, which features solitonic solutions[23]. We observe linear and nonlinear beam propagation, for which the later one shows a clear localization. Besides resting solitons, we observe also self-accelerating nonlinear solution in 2D.

**Results**

**Experimental setup and theoretical model.** Our experimental platform (see Fig. 1a) is based on four slightly dissimilar coupled fiber loops of a length of approximately 30 km. They are grouped in two pairs, each standing for one synthetic transverse dimension (see Supplementary Note 2), as demonstrated for 1D[28] and 2D[29] lattices. The two inner loops A and B differ by $\Delta L_{\text{inner}} = L_A - L_B \approx$ 600 m ($\Delta T_{\text{inner}} = 3$ μs), while the two outer loops C and D differ by $\Delta L_{\text{outer}} = L_C - L_D \approx 6$ m ($\Delta T_{\text{outer}} = 30$ ns). As shown in Fig. 1a, an initial seed pulse is injected via a fiber optical coupler into the outer left loop C and splits into two pulses (step Ia and Ib) at the first 50/50 coupler at the entrance of the two inner loops A and B. After passing through the second 50/50 coupler, the pulses split again (step IIa and IIb) and propagate as pairs through the outer loops C and D. They return with varying delay at the first 50/50 coupler after a mean round trip time $\overline{T} \approx 300$ μs. Here the journey starts again. After $m$ round trips a pulse sequence has formed, in which each pulse arrives at a time

$$T_{\text{arrival}}(m, x, y) = m\overline{T} + x\,\Delta T_{\text{inner}}/2 + y\,\Delta T_{\text{outer}}/2,\tag{1}$$

where the integer numbers $x$ and $y$ denote how more often the pulse has passed the longer than the shorter inner and outer loops, respectively. Because $x$ and $y$ are always smaller than $m$ and as long as $m\,\Delta T_{\text{outer}} < \Delta T_{\text{inner}}$ and $m\,\Delta T_{\text{inner}} < \overline{T}$ holds, $x$ and $y$ are uniquely determined by the arrival time. In our case this allows for $m<100$ for a straightforward mapping onto an equivalent 2D mesh lattice spanned by $x$ and $y$ (see Fig. 1b,d). By passing loop A (B), $x$ increases (decreases) by one, which is equivalent to a step to the right (left) on the 2D synthetic lattice. Afterwards, by propagating through the outer loop C (D), $y$ increases (decreases) by one, corresponding to a step up (down) on the lattice. In this way, any path through the 2D lattice is equivalent to a combination of roundtrips through the four different loops (see the pathways in Fig. 1b) and vice versa. The pulse sequence evolving in the system is measured by photo detectors (blue curve in Fig. 1d), sampled electronically (red dotted curve in Fig. 1d), and mapped onto a 2D discrete lattice in x and y according to Eq.1 (see insets in Fig. 1d for time steps $m = 1,2$ and 3). As we use 22 ns long pulses, pulse dispersion is negligible and the overall measurement of photodetected pulses are evaluated by averaging a measurement slot ($M_{slot} \approx m_{max}\overline{T}$, where $m_{max}$ is the maximum number of time steps) over 100 times. Altogether, the maximum size of the resembled 2D synthetic lattice is defined as $x_{\max} \approx \overline{T}/\Delta T_{\text{inner}}$ and $y_{\max} \approx \Delta T_{\text{inner}}/\Delta T_{\text{outer}}$ in order to prevent those pulses of adjacent roundtrips overlap.

Numerically, the overall dynamics is well described by complex field amplitudes $a_{x,y}^m$ / $b_{x,y}^m$ and $c_{x,y}^m$ / $d_{x,y}^m$ for pulses traveling through A and B loop (short / long inner loops) and C and D loop (short / long outer loops), respectively. By interpreting the number of roundtrips $m$ as the discrete time and $(x,y)$ as the position on the synthetic lattice, the pulse evolution is described by the evolution equations for the inner loops as

$$a_{x,y}^m = \sqrt{\frac{G_a(m)}{2}}\left(c_{x+1,y}^{m-1} + id_{x+1,y}^{m-1}\right)\exp\left(i\varphi_a(x,y) + i\chi\left|c_{x+1,y}^{m-1} + id_{x+1,y}^{m-1}\right|^2\right) \text{ and}\tag{2}$$

$$b_{x,y}^m = \sqrt{\frac{G_b(m)}{2}}\left(d_{x-1,y}^{m-1} + ic_{x-1,y}^{m-1}\right)\exp\left(i\varphi_b(x,y) + i\chi\left|d_{x-1,y}^{m-1} + ic_{x-1,y}^{m-1}\right|^2\right),\tag{3}$$

and for the outer loops as

$$c_{x,y}^m = \sqrt{\frac{G_c(m)}{2}}(a_{x,y+1}^m + ib_{x,y+1}^m)\exp\left(i\varphi_c(x,y) + i\chi|a_{x,y+1}^m + ib_{x,y+1}^m|^2\right) \text{ and} \quad (4)$$

$$d_{x,y}^m = \sqrt{\frac{G_d(m)}{2}}(b_{x,y-1}^m + ia_{x,y-1}^m)\exp\left(i\varphi_d(x,y) + i\chi|b_{x,y-1}^m + ia_{x,y-1}^m|^2\right), \quad (5)$$

where $G_{a,b,c,d}$ stands for the adjustable net gain ($G$) and loss ($1/G$) introduced by the amplitude modulators (AM) in each fiber loop, where it exchanges gain and loss after every round trip (see Method Section 1 and Fig. 1c). The second part of Eqs. (2)-(5) represents the interference of previous pulses from their adjacent lattice sites inside the 50/50 coupler. Additionally, the third part denotes a combination of linear and nonlinear phase increment by, respectively, phase modulators (PM) and power-dependent nonlinear phase shift[30] ($\varphi_{NL} = e^{i\chi|a|^2}$) proportional to the effective fiber nonlinearity $\chi$ (see Supplementary Note 7). Specifically, PMs apply a phase pattern $\varphi_a$, $\varphi_b$, $\varphi_c$, and $\varphi_d$ to the pulses depending on their position on the lattice (see Method Section 2) in each time step $m$ in order to fulfill 2D $\mathcal{PT}$-symmetry[31] (see Fig. 1b and Supplementary Note 9). In this model, the ease alternation of gain/loss factor each time step $m$ is just achievable since the AMs are placed with an idle transmission ratio of 0.80 in the passive case, which the pulses can be easily amplified (or attenuated) by setting a higher (or smaller) value of transmission ratio. Importantly, in order to compensate all signal losses caused by the idle transmission ratio, absorption and monitoring of the optical components, an erbium-based fiber amplifier (EDFA) is used in each loop for restoring the energy conservation of the system and enabling a considerable increase of propagation steps[28].

**Broken and recovered $\mathcal{PT}$-symmetry regions.** In the linear case ($\chi = 0$), the band structure

$$\cos\theta = \pm\frac{1}{8}\left(-2\cos(g) + \cos(k_x - k_y) - 4\cos(\varphi_o)\sin^2\left(\frac{k_x + k_y}{2}\right)\right.$$
$$\pm\sqrt{2}\cos\left(\frac{k_x + k_y}{2}\right)[14 - 6\cos(2\varphi_o) + 4\cos(\varphi_o - g)$$
$$+4\cos(\varphi_o + g) + \cos(2\varphi_o - k_x - k_y) + 4\cos(\varphi_o + k_x - k_y)$$
$$+4\cos(\varphi_o - k_x + k_y) + 4\cos(g - k_x + k_y) + 4\cos(g + k_x - k_y)$$
$$\left.-2\cos(k_x + k_y) + \cos(2\varphi_o + k_x + k_y)]^{1/2}\right). \quad (6)$$

of the system is calculated by inserting the evolution equations (2)-(5) into a Floquet-Bloch ansatz of the form[32]

$$U_{\text{PT}_{x,y}}(g,\varphi_0)e^{(i\theta m - i(k_x x + k_y y))} \begin{pmatrix} A_1 \\ B_1 \\ A_2 \\ B_2 \end{pmatrix}_{k_x,k_y} = e^{(i\theta(m+1) - i(k_x x + k_y y))} \begin{pmatrix} A_1 \\ B_1 \\ A_2 \\ B_2 \end{pmatrix}_{k_x,k_y} \quad (7)$$

where $U_{\text{PT}_{x,y}}$ is the evolution operator in presence of $\mathcal{PT}$-symmetric potentials. The phase and amplitude modulation are denoted by $\varphi_o$ and $g = -2i\,ln(G)$, respectively. $\theta$ stands for the propagation constant and $k_x$ and $k_y$ are the quasi momenta (see Supplementary Note 6, 8). The Bloch states are given by the double-step two-component vector $(A_1, B_1, A_2, B_2)_{k_x,k_y}^{\text{T}}$, which represents the amplitude and phase relation between the shorter and longer loops. The following eigenvector problem

$$U_{\text{PT}_{x,y}}(g,\varphi_0) \begin{pmatrix} A_1 \\ B_1 \\ A_2 \\ B_2 \end{pmatrix}_{k_x,k_y} = e^{i\theta} \begin{pmatrix} A_1 \\ B_1 \\ A_2 \\ B_2 \end{pmatrix}_{k_x,k_y} \quad (8)$$

and its corresponding eigenvalues $\lambda$ delivers four quasi-energy bands $\theta = -i\ln\left(\lambda(k_x, k_y)\right)$. Accordingly to the 2D $\mathcal{PT}$-symmetry, an antisymmetric gain/loss modulation is required[33], which is here implemented by amplifying and attenuating the shorter/longer inner and outer loops in a balanced way, where gain and loss are swapped every roundtrip (see Fig. 1b). This creates plaques of gain and loss (see Fig. 1c). Similarly, the simplest phase modulation that satisfies the 2D $\mathcal{PT}$-symmetry condition[33] has a spatial periodicity of four positions (see Fig. 1b and Supplementary Note 6). However, this doubles the unit cell of the lattice and thus the two original bands split into four in total. Interestingly, the $\mathcal{PT}$-symmetric phase modulation depicted in Fig. 1b creates zigzag-shaped potential barriers along the lattice, similarly to the effective Peierls-Nabarro (PN) barrier[34].

By inserting a single pulse in the C loop, which represent a single excitation onto the center of lattice ($x = y = 0$), the entire band structure is excited in the momentum space, including the upper and the lower band. Consequently, the system has non-imaginary components (see Fig. 2a) for the passive and conservative case ($G = 1.0$, $\varphi_a, = \varphi_b = \varphi_c = \varphi_d = 0$) and thus performs a 2D light walk (see Fig. 3a). However, for $G > 1.0$ and without any phase modulation, the band structure presents complex values ($Im(\theta) > 0$) and $\mathcal{PT}$-symmetry is broken (see Fig. 2b,c), thus leading to a power that grows exponentially during the propagation, as shown in Fig. 3b,c for $\varphi_0 = 0$ and $0.3\pi$. In order to restore a pseudo-Hermitian evolution, a symmetric phase potential is applied in combination with the gain and loss modulation, so that $\mathcal{PT}$-symmetry is fulfilled (see Fig. 2d and Fig. 3b,c for $\varphi_0 = 0.6\pi$). In presence of this phase modulation, the energy is conserved on average during propagation (see Fig. 3d), which is consistent with the real-valued band structure in Fig. 2d. Similar to the 1D case[17,18], we find the gain/loss factor threshold at which the 2D $\mathcal{PT}$ phase can recover its real-valued and quasi-conservative dynamics, as shown in Fig. 2 (lower chart).

**Nonlinear propagation in 2D quasi-Hermitian synthetic lattice.** Considering the recovered $\mathcal{PT}$-symmetry case ($G = 1.1$, $\varphi_0 = 0.6\pi$), which presents a bandgap and real-valued band structure ($Im(\theta) = 0$), the upper band approaches the dispersion relation of waves propagating in bulk materials since it has a constant positive curvature in a wide momentum range. In contrast to a single lattice site excitation (a single pulse), which populates the entire band structure in the momentum space, a specific point-like region of the Brillouin zone can be excited by a wave packet that is broad in the real space. In this configuration, a broad excitation containing a selective population within the central point $\Gamma$ ($k_x = k_y = 0$) of the upper band is carried out by launching a train of rectangular pulses having a Gaussian envelope $G_w(x, y) = A_w \exp[-(x^2 + y^2)/w^2]$ along $x$- and $y$-axis with a variable amplitude ($A_w$) and a fixed width ($w$) of 6 positions ($1/e$ drop of intensity) (see Fig. 4a and for more details of the preparation see Supplementary Note 10).

At input power of approximately 0.208 mW, the field distribution experiences diffraction and spreads linearly on the synthetic lattice, which follows a diagonal spreading in the $x$-$y$-plane (see Figs. 4c,d,e) due to the orientation of the phase potential lines (as shown in Fig. 1b). By accurately increasing the input power at 1.1 mW, 2D $\mathcal{PT}$ solitons are created (see Fig. 4f,g,h). Numerical simulations (see Method Section 3) show that the soliton lifetime highly depends on the soliton total energy $E = \sum_{x,y=1}^{N}\left(\left|a_{x,y}\right|^2 + \left|b_{x,y}\right|^2\right)$ and gain/loss factor and, consequently, the nonlinear solutions either broaden or compress for propagation distances that are inaccessible to the experiment. Interestingly, likewise in 1D $\mathcal{PT}$ mesh lattices[18,35], the solitons also feature a one-parametric family. In this way, the $\mathcal{PT}$ system mimics its Hermitian counterpart and allows the solitons to adapt their amplitudes to their widths. Also, similar to the Townes-like soliton of the conservative 2D nonlinear Schrödinger equation[36,37], the 2D $\mathcal{PT}$ solitonic waves are intrinsically unstable. The variation of the soliton propagation constant (eigenvalue $\theta$) for the conservative ($G = 1.0$) and non-Hermitian systems ($1.01 \leq G \leq 1.76$), as a function of the total energy ($E$), are depicted in Fig. 5a. The intensity profile of low energetic solutions is considerably asymmetric with respect to diagonal directions (compare Fig. 4e and h). This agrees with the quasi one-dimensional pattern of that $\mathcal{PT}$ phase, which can be invariantly interpreted (with a discrete step) as a tooth-like potential along the $x = y$ direction. In contrast, high energetic solutions appear more symmetric in shape when their width almost approaches one elementary $\mathcal{PT}$ unit cell (see Fig. 4h), thus corresponding to a highly localized soliton trapped between two zigzag-shaped phase potential barriers (as a PN barrier[34]). Similarly, Fig. 5b shows the eigenvalue–soliton width curve, where the diagonal $x = y$ is considered for fitting a Gaussian field distribution. Accordingly to these figures, as the total power increases, soliton eigenvalues move further into the band gap. Interestingly, the conservative soliton line (dotted black line) determines the threshold of the propagation constant at which non-conservative nonlinear localized stationary solutions (i.e. $G > 1.0$) cannot exist. As the gain/loss factor gets increased, the corresponding propagation constant curves for $\mathcal{PT}$ solitons proportionally decrease and their widths rapidly become narrower. Also, non-conservative soliton eigenvalues present a total energy that is limited approximately to 1.7 due to their higher dissipative flux of energy for bigger gain/loss factors (see Fig. 5c), which make them more unstable and lead rapidly either a blow-up or collapse event.

Although unstable but stationary solutions, broad low energetic solitons ($w \approx 5$ unit cells) are noticeably larger than the step size of the lattice and the size of the potential zigzag corners ($\mathcal{PT}$-symmetric phase modulation). As a result, discretization effect of the lattice and effective PN barrier become negligible in this continuous limit, thus allowing the soliton to live extremely long propagation times (see Fig. 5d). Since the PN potential is diminished, the broad solitons are able to diffract at the corners and very slowly move along the diagonal zigzag-shaped potential without any energy loss[26]. On the other hand, in Fig. 5d, higher energetic, non-Hermitian solitons are more stable for longer time steps $m$ as long as gain/loss factor is smaller. It is found that, despite the waves propagating in a recovered $\mathcal{PT}$-symmetry case, the soliton energy flux inherently exhibits a small effective energy growth factor proportional to the gain factor ($G$). For any $G > 1$, the $\mathcal{PT}$ soliton sweeps along the characteristic curve and gets narrower with time and, consequently, it suggests that $\mathcal{PT}$-symmetry is locally broken by the nonlinear solution. As non-Hermitian solitons propagate on the quasi-conserved system, they do not get immediately destroyed, but instead its energy exponentially grows until the region of instability is reached and thus leading to the collapse of the soliton (see Supplementary Note 11). Altogether, the initial unstable soliton ($\theta \gtrsim -0.22\pi$) does not turn into a less energetic solution, but instead it is almost adiabatically collapsed into a highly localized non-stationary state, which demonstrate the typical process for the conservative 2D discrete Schrödinger system[38,39]. Since the transition under consideration makes the soliton abruptly shrink in space, we can similarly refer to it as a collapse event on the discrete lattice.

When choosing higher input power, for instance 4.15 mW in the experiment, the region of instability is reached faster and the nonlinear self-focusing leads to a collapse of the field distribution (see Figs. 4i). In contrast to its conservative counterpart ($G = 0$), the non-Hermitian collapse event is followed by a faster growth of the total energy (see red curve in Fig. 4b), which shows a stronger local break of the $\mathcal{PT}$-symmetry. The extremely localized field is concentrated around a single lattice site and a small amount of excess radiation is released in the form of moving outwards free propagating waves (see Figs. 4i,j). At the collapsing event, the highly localized wave nonlinearly self-accelerates and moves on the lattice (see Figs. 4k). By numerical investigation, the directionality can be presumed and tends to be, in the most cases, perpendicular to the zigzag-shaped $\mathcal{PT}$ phase potentials (PN barrier). This fact suggests that the localized energy flux of the initial stationary non-Hermitian soliton are oriented, in total, along the diagonal direction, in accordance with the $\mathcal{PT}$-symmetric nodes of energy sources (gain) and drains (loss). Nevertheless, due to a very small width ($w \approx 1$) of the moving localization, it experiences discretization effect of the lattice as well as the zigzag-shaped $\mathcal{PT}$ phase potentials (see Supplementary Note 11). Consequently, as a result of overcoming the phase barrier, the highly localized, collapsed soliton while moving gradually lose its energy until it drops below a certain threshold, making it dissolve.

In conclusion, we successfully realized a novel $\mathcal{PT}$-symmetric system in a 2D synthetic lattice based on telecomm equipment. By tuning gain/loss and phase modulation, we observe a single pulse evolution in the pseudo-Hermitian case within the unbroken $\mathcal{PT}$-symmetry, while unstable one for broken regimes. Additionally, we accomplished non-Hermitian nonlinear localization of a broad Gaussian-like initial field distribution by exploiting eigenvalue spectrum separated by a bandgap. In contrast to the Hermitian case, non-conservative $\mathcal{PT}$ solitons present an effective energy growth, which make them more unstable and lead rapidly either a blow-up or collapse event. For higher power levels, a family of non-Hermitian solitons is investigated, which its nonlinear instability solutions lead to a collapse event and create a self-accelerating and nonlinearly-driven motion of the field distribution.

**Methods**

**Method 1: $\mathcal{PT}$-symmetric gain/loss modulation.** In order to satisfy $\mathcal{PT}$-symmetry condition, the gain/loss pattern (imaginary part of the potential) should be antisymmetric with respect to central symmetry point $x = -y$ (see Fig. 1b). Therefore, a gain/loss modulation provided by AMs applies an amplitude difference on each lattice arm, resembling gain ($G$) and loss ($1/G$) factors. As AMs are on idle transmission ratio of 0.8 in the passive case, a maximum achievable gain factor for one time step is $1/0.8 = 1.25$. The final gain/loss pattern, which is just dependent on time step $m$, for A and B loop can be written as

$$\begin{array}{ll} G_a(m) = G & \text{if } \mathrm{mod}(m,2) = 0 \\ G_b(m) = 1/G & \text{if } \mathrm{mod}(m,2) = 1 \end{array},$$

following for C and D loop as

$$\begin{array}{ll} G_c(m) = 1/G & \text{if } \mathrm{mod}(m,2) = 0 \\ G_d(m) = G & \text{if } \mathrm{mod}(m,2) = 1 \end{array}.$$

**Method 2: $\mathcal{PT}$-symmetric phase modulation.** $\mathcal{PT}$-symmetry condition for phase modulation (real part of the potential) should be symmetric with respect to central symmetry point $x = -y$. In contrast to gain/loss modulation, the phase modulation applies an alternating phase potential dependent on the time step $m$ and position $x$ and $y$ on the lattice (see Fig. 1b). For even time steps, its pattern thus reads for loop A and B an opposite alternation (0 or $\varphi_0$), whereas loop C and D comprise of the same alternating pattern as loop A, as shown in Fig. 1b. For odd time steps, all phase modulations are set to zero. Altogether, $\mathcal{PT}$ phase modulation for even time steps $m$ can be summarized for A, C and D loops as

$$\begin{aligned}\varphi_{a,c,d}(x,y) &= \varphi_0, &&\text{if } \mathrm{mod}(x,4) = 0 \text{ and } \mathrm{mod}(y,4) = 0 \\ \varphi_{a,c,d}(x,y) &= \varphi_0, &&\text{if } \mathrm{mod}(x,4) = 2 \text{ and } \mathrm{mod}(y,4) = 2 \;, \\ \varphi_{a,c,d}(x,y) &= 0, &&\text{otherwise}\end{aligned}$$

and for B loop

$$\begin{aligned}\varphi_b(x,y) &= \varphi_0, &&\text{if } \mathrm{mod}(x,4) = 2 \text{ and } \mathrm{mod}(y,4) = 0 \\ \varphi_b(x,y) &= \varphi_0, &&\text{if } \mathrm{mod}(x,4) = 0 \text{ and } \mathrm{mod}(y,4) = 2 \;, \\ \varphi_b(x,y) &= 0, &&\text{otherwise}\end{aligned}$$

whereas for odd time steps $m$, they follow as

$$\varphi_{a,b,c,d}(x,y) = 0\,.$$

**Method 3: Numerical investigation of nonlinear solutions.** We investigate the nonlinear localized stationary solutions of Eqs. (2)-(5) for the case $\varphi_0 = 0.6\pi$ and $G \geq 1$. The nonlinear coefficient $\chi$ is fixed to 1, so that the localized solutions originate from the positive ($\theta > 0$) focusing band. The total energy

$$E = \sum_{x,y=1}^{N} \left( |a_{x,y}|^2 + |b_{x,y}|^2 \right)$$

is unconstrained by the solver as it can be freely varied during the optimization process. The domain has periodic boundaries in both x and y directions. The size of the computational domain N is $80 \times 80$ positions, which corresponds to $40 \times 40$ elementary unit cells of the $\mathcal{PT}$-symmetric lattice. The optimization process is based on the in-built Matlab Levenberg-Marquardt algorithm, which aims to minimize the following nonlinear multidimensional problem:

$$\begin{cases} \left| a_{x,y}^{m=2} - a_{x,y}^{m=0} e^{i\theta} \right|^2 \to 0 \\ \left| b_{x,y}^{m=2} - b_{x,y}^{m=0} e^{i\theta} \right|^2 \to 0 \end{cases},$$

where $\theta$ stands the propagation constant and the originally two-dimensional vector $\{a_{x,y}, b_{x,y}\}_{x,y=1...N}$ is preliminary stacked into one-dimensional one of the form $\{a_{1,1}, b_{1,1}, a_{1,2}, b_{1,2} \dots a_{1,N}, b_{1,N}, a_{2,1}, b_{2,1} \dots a_{N,N}, b_{N,N}\}$ and the double step propagator for such a state is a consecutive action of a matrix $N^2 \times N^2$ (linear operations) and that of the nonlinear phase shift, depending on the amplitude distribution of the state itself, as follows from the evolution equations in Eqs. (2)-(5). Function tolerance of the algorithm was typically set to $10^{-7}$, however spatially narrower solutions were found to gradually destabilize at higher $\theta$ and thus the required precision for them had to be lowered down to $10^{-5}$ in order for the algorithm converge in a manageable time. As the initial trial function, we choose a radially symmetric Gaussian envelope of 10 position width and the Bloch eigenvector of the elementary unit cell, that corresponds to the central point ($k_x = k_y = 0$, $\theta_{\text{band}} > 0$) of the upper focusing band. At the beginning, the propagation constant ($\theta$) of the target solution is fixed slightly above $\theta_{\text{band}}$ of the corresponding linear Bloch wave. After the solution is found, it is further chosen as a trial function for the next target solution, whose propagation constant is again slightly increased with respect to the previous one. In this way, the parametric family of solitons can be traced up to almost $-\theta_{\text{band}}$, where the lower linear focusing band is closing the gap (see Supplementary Note 11).

doi:10.1103/PhysRevA.85.063837

36. Moll, K. D., Gaeta, A. L. & Fibich, G. Self-Similar Optical Wave Collapse: Observation of the Townes Profile. *Phys. Rev. Lett.* (2003). doi:10.1103/PhysRevLett.90.203902

37. Sulem, C. & Sulem, P. L. The nonlinear Schrödinger equation : self-focusing and wave collapse. *Appl. Math. Sci.* (1999). doi:10.1007/b98958

38. Szameit, A. *et al.* Two-dimensional soliton in cubic fs laser written waveguide arrays in fused silica. *Opt. Express* (2006). doi:10.1364/OE.14.006055

39. Aceves, A. B., Luther, G. G., De Angelis, C., Rubenchik, A. M. & Turitsyn, S. K. Energy localization in nonlinear fiber arrays: Collapse-effect compressor. *Phys. Rev. Lett.* (1995). doi:10.1103/PhysRevLett.75.73

**Acknowledgments**

This project was supported by the German Research Foundation (DFG) through the International Research Training Group (IRTG) 2101, as well as an NSERC CREATE grant.

# Figures

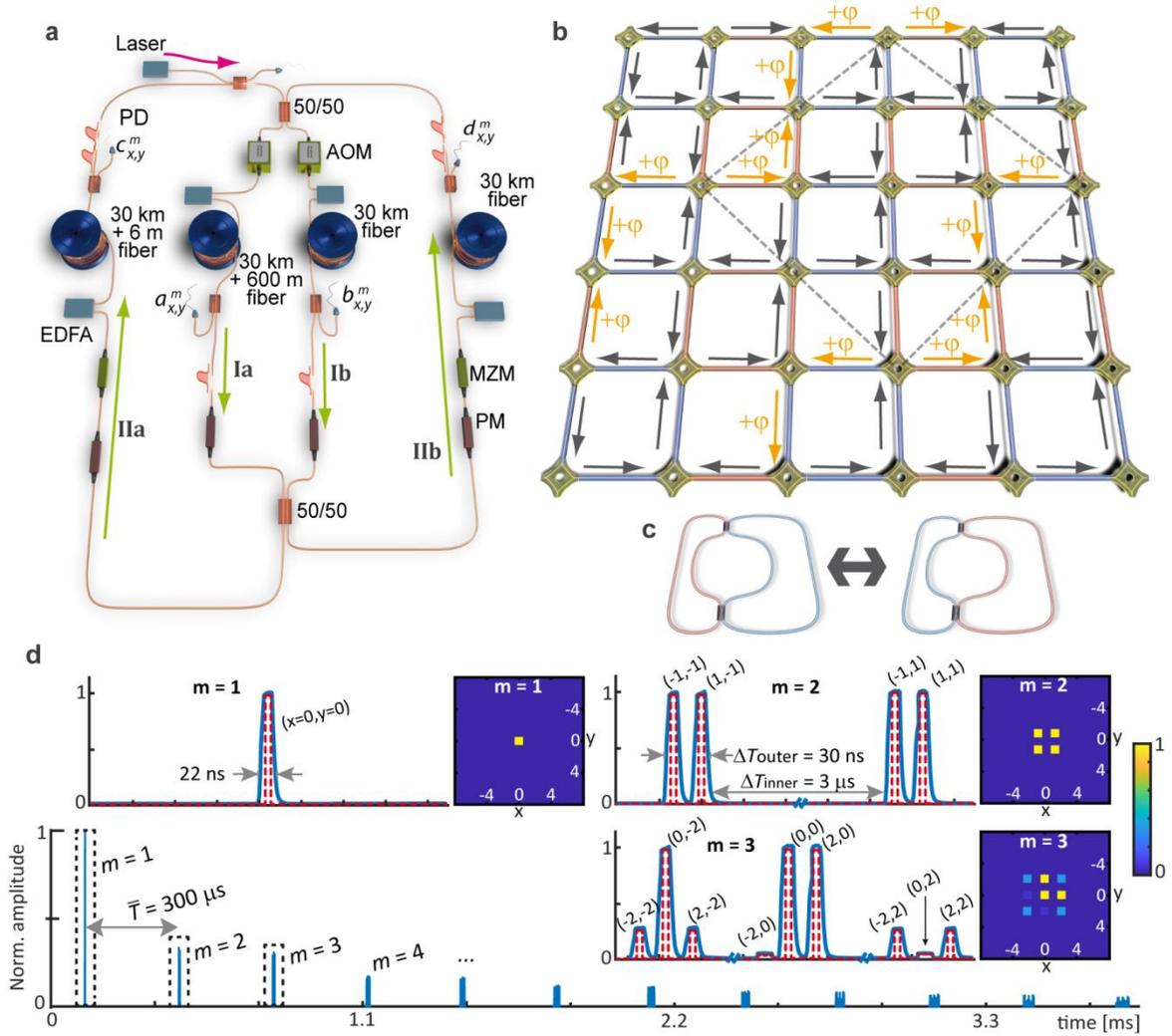

**Figure 1. Light propagation on a 2D mesh lattice. a**, The inner ($a_{x,y}^m$ and $b_{x,y}^m$) and outer ($c_{x,y}^m$ and $d_{x,y}^m$) pair of fibers are connected via 50/50 couplers. A pulse is created and injected into the outer left loop. Each fiber path has an erbium-doped fiber amplifier (EDFA) for loss compensation and 30 km of optical fiber. Acoustic optical modulators (AOM) in the inner and Mach-Zehnder modulators (MZM) in the outer pair allow for amplitude modulation. Additionally, phase modulators (PM) are placed in the both loops. **b**, The $\mathcal{PT}$–symmetric 2D synthetic lattice is virtually mapped accordingly to the arrival pulses in each loop, considering loop A (horizontal arrows pointing to the right), loop B (horizontal arrows pointing to the left), loop C (vertical arrows pointing to upward) and loop D (horizontal arrows pointing to downward). Accordingly, dashed grey lines represent the $\mathcal{PT}$ unit cell; red and blue lattice arms display gain and loss, respectively; and $\varphi$ denotes a phase modulation. **c**, Varying proportionally the transmission ratio of those amplitude modulators in each loop every coupling lengths $m$ creates plaques of gain (red) and loss (blue) in order to fulfill $\mathcal{PT}$–symmetry. **d**, During each round trip, each pulse splits and interferes into the 50/50 couplers and arrives with different arrival times at the photodetectors, which their amplitudes are measured by their photodetected electrical power (blue line), sampled electronically (red dashed line) by a computer software and mapped onto a 2D spatially $x$-$y$ representation.

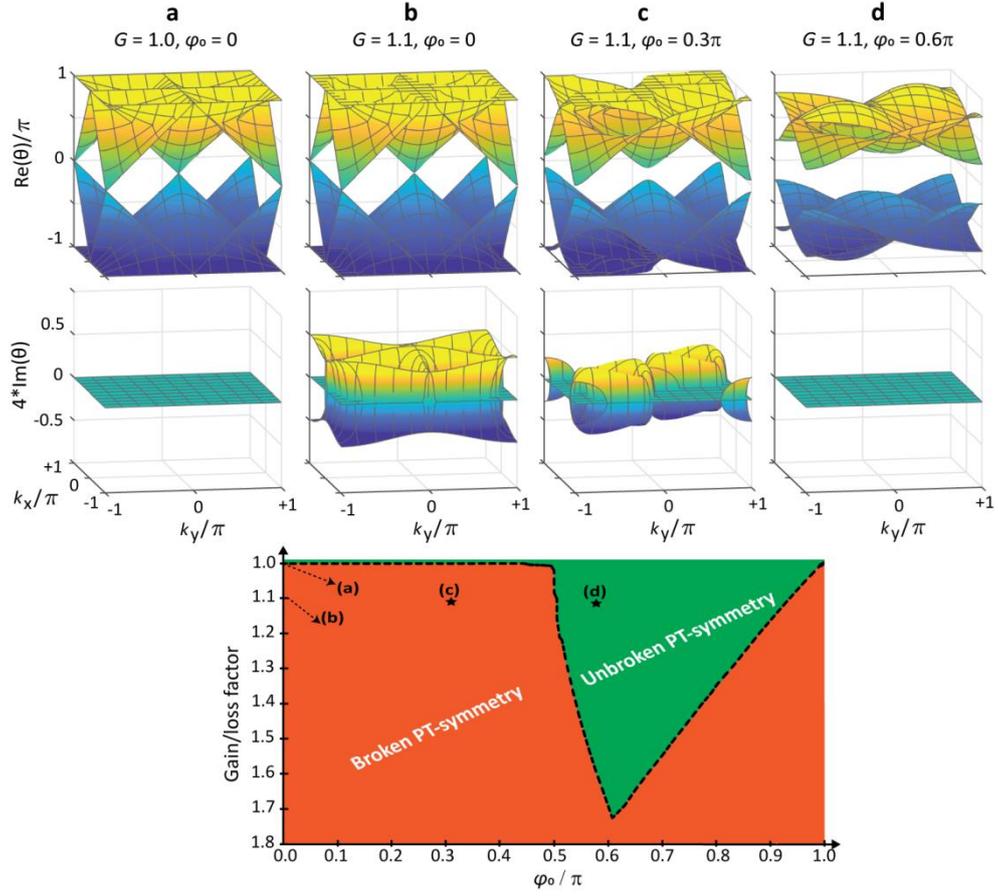

**Figure 2. Band structure of the 2D $\mathcal{PT}$-symmetric mesh lattice and quasi-conservative regions.** Due to the Floquet nature of the system, not only the Bloch momenta $k_{x/y}$, but also the propagation constant $\theta$ is periodic within $[-\pi; \pi]$. **a,** Passive band structure in absence of any gain/loss and phase modulation. **b-d,** Band structure in presence of gain/loss of $G = 1.1$ and phase potential $\varphi_0$ of **(b)** $0$, **(c)** $0.3\pi$, **(d)** $0.6\pi$. Lower figure represent the broken (orange) and recovered (green) $\mathcal{PT}$-symmetry region as a function of gain/loss and phase potentials.

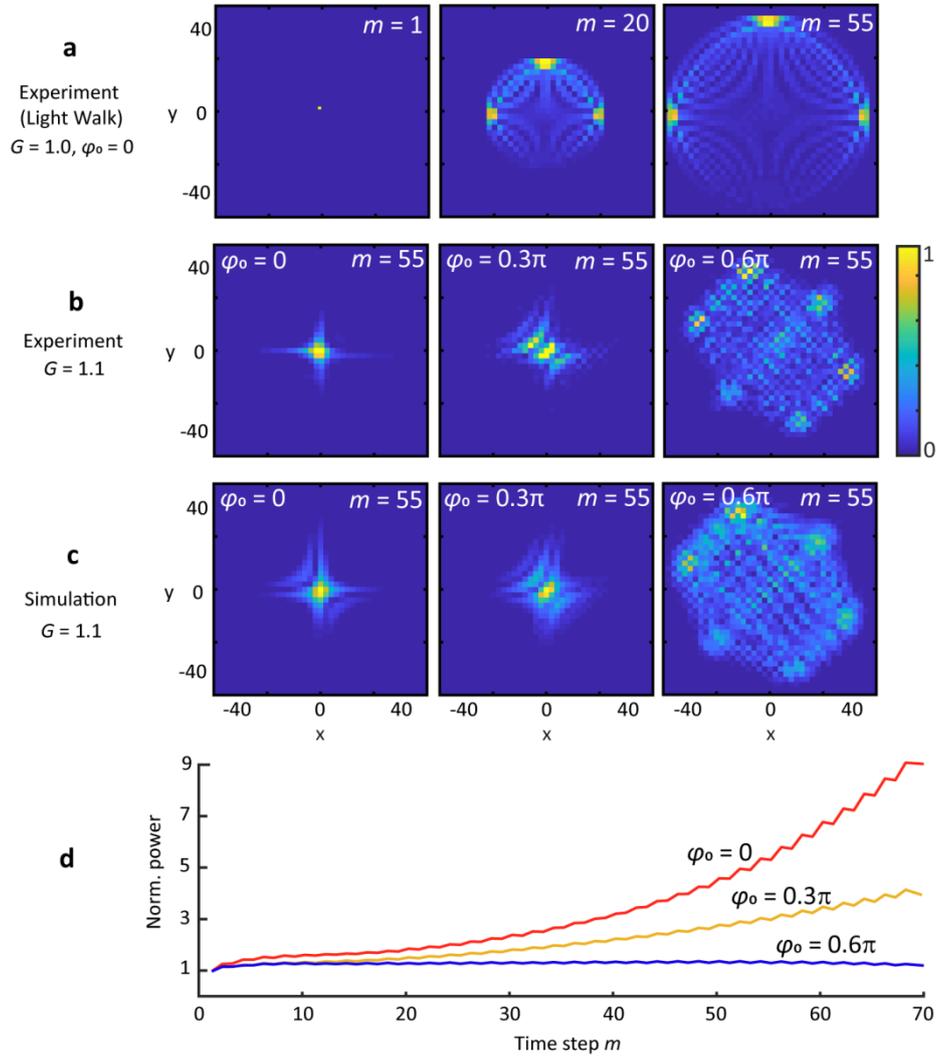

**Figure 3. Evolution of a single site excitation in the presence of $\mathcal{PT}$-symmetric potentials. a**, different propagation steps $m$ of an experimentally realized 2D Light walk for vanishing phase modulation $\varphi_0 = 0$ and passive case $G = 0$. **b**, experiment and **c**, simulation of a single site excitation at $m = 55$ in presence of gain/loss $G = 1.1$ and phase modulation $\varphi_0 = 0, 0.3\pi$ and $0.6\pi$. The latter one recovers a pseudo-Hermitian evolution and $\mathcal{PT}$-symmetry is fulfilled. **d**, Experimental investigation of the evolution energy as a function of coupling lengths $m$ for $G = 1.1$ and phase modulation $\varphi_0 = 0, 0.3\pi$ and $0.6\pi$.

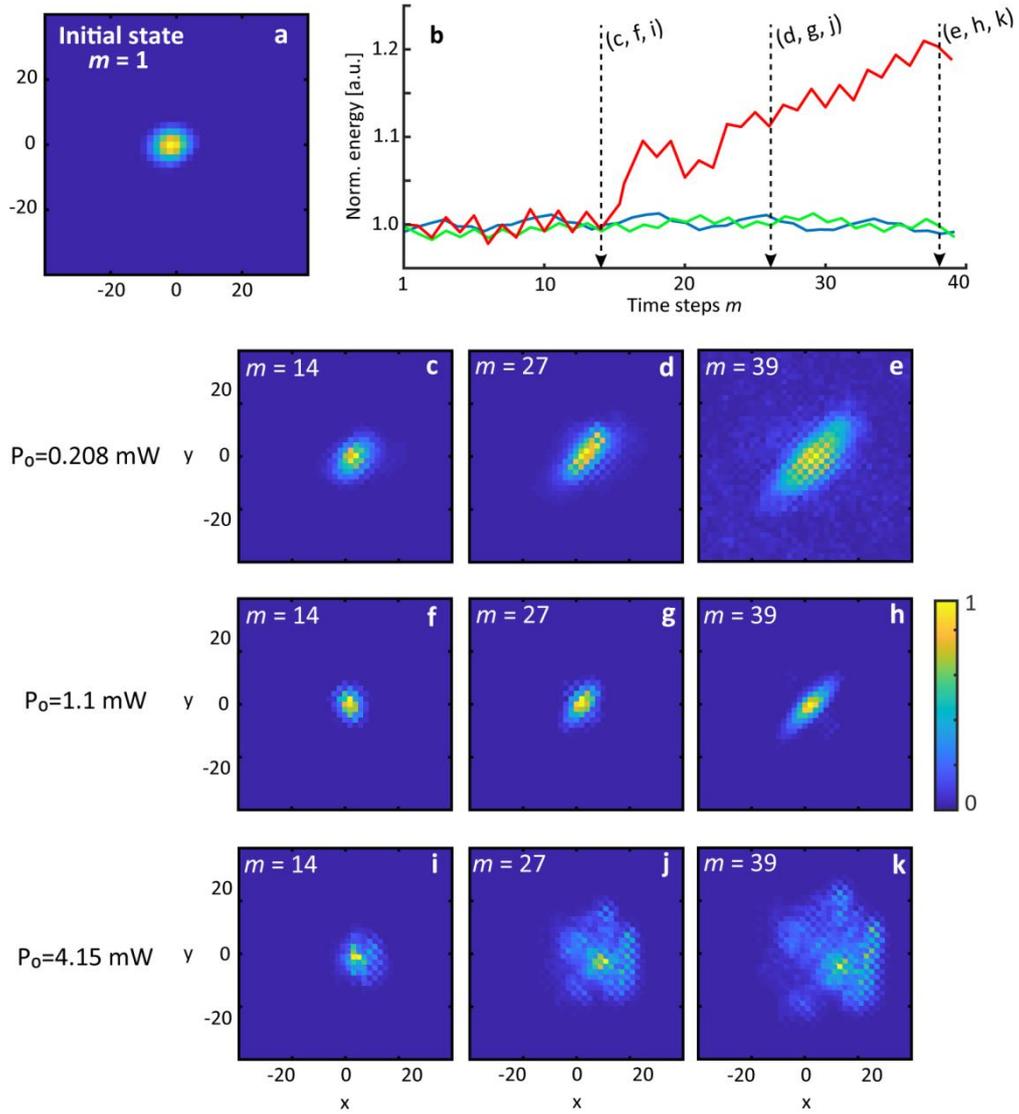

**Figure 4. Evolution of a broad excitation in the presence of $\mathcal{PT}$–symmetric potentials ($G = 1.1$, $\varphi_0 = 0.6\pi$) and for different power levels demonstrating soliton formation and wave collapse at the highest power level. a**, 2D image of the initial Gaussian distribution after its preparation. **b**, Experimental investigation of the evolution energy as a function of time steps $m$ for 0.208 (blue), 1.1 (green), 4.15 mW (red). **c-k**, 2D image displayed with normalized scaled colors of the wave packets after 14 (**c,f,i**), 27 (**d,g,j**) and 39 (**e,h,k**) time steps $m$ for different input powers (0.208, 1.1 and 4.15 mW).

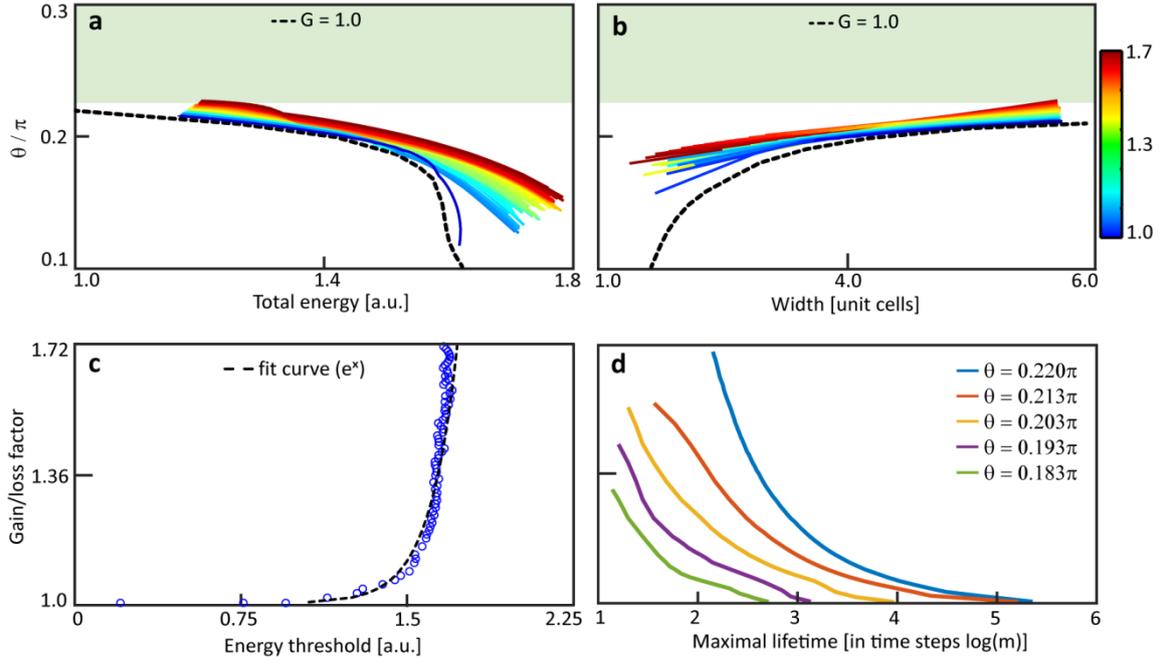

**Figure 5. Numerical simulations of the soliton in the conservative and $\mathcal{PT}$–symmetry systems.** The values for the propagation constant of the soliton as a function of total energy and width are lying inside the band gap. **a**, From the upper edge of the band gap, soliton solutions move into the gap as the total energy $E$ increases for conservative solitons (dotted curve, $G = 1.0$ and $\varphi_0 = 0.6\pi$), whereas non-Hermitian solitons (from $G = 1.01$ to $G = 1.76$, and $\varphi_0 = 0.6\pi$) exists only in a small span of total energy depending on its gain/loss potential (from red to blue scale). **b**, The width $w$ of the conservative soliton (dotted curve, $G = 1.0$ and $\varphi_0 = 0.6\pi$), at the edge of the band gap, tends to a large size and it decreases to a minimum width ~1 positions distant from the edge. Non-conservative solitons display a limited width positions until they either blow-up or collapse. **c**, Total energy threshold and **d**, maximal lifetime (in time step $m$) as a function of gain/loss potential.